\documentclass[12pt]{article}
\usepackage{a4}
\usepackage{epsfig}
\begin{document}
\begin{center}
{\bf THE SPONTANEOUS GENERATION OF MAGNETIC FIELDS AT HIGH TEMPERATURE
IN A SUPERSYMMETRIC THEORY}
                                ~\\
                                ~\\

 { Vadim Demchik$^a$ and Vladimir Skalozub$^a$ } \\
                                  ~\\
                                   ~\\
$^a$ Dniepropetrovsk National University, 49050 Dniepropetrovsk, Ukraine\\
e-mail:  vadimdi@yahoo.com; Skalozub@ff.dsu.dp.ua
\\
                         ~\\
                        ~\\
                       {\bf ABSTRACT}
\end{center}
The spontaneous generation of magnetic and chromomagnetic fields
at high temperature in the minimal supersymmetric standard model
(MSSM) is investigated. The consistent effective potential
including the one-loop and the daisy diagrams of all bosons and
fermions is calculated and the magnetization of the vacuum is
observed. The mixing of the generated fields due to the quark and
s-quark loop diagrams and the role of superpartners are studied in
detail. It is found that the contribution of these diagrams
increases the magnetic and chromomagnetic field strengths as
compared with the case of a separate generation of fields. The
magnetized vacuum state is found to be stable due to the magnetic
masses of gauge fields included in the daisy diagrams.
Applications of the results obtained are discussed. A comparison
with the standard model case is done.

\subsection*{Introduction}
Possible existence of strong magnetic fields in the early universe
is one of the most interesting problems in high energy physics.
Different mechanisms of the fields producing at different stages
of the universe evolution were proposed. These mechanisms as well
as the role of strong magnetic fields have been discussed in the
surveys \cite{Enq}-\cite{Grasso}. In particular, primordial
magnetic fields, being implemented in a cosmic plasma, may serve
as the seed ones for the present extra-galaxy fields.

The spontaneous vacuum magnetization at high temperature is one of
the mechanisms mentioned. It was already investigated both in
pure $SU(2)$ gluodynamics \cite{Enq3}-\cite{Once} and in the
standard model (SM) \cite{DS} where the possibility of this
phenomenon has been shown. The stability of the magnetized vacuum
state was also studied \cite{Once}. The magnetization takes place
for the non-abelian gauge fields due to a vacuum dynamics. In
fact, this is one of the distinguishable features of
asymptotically free theories \cite{Once,Sav}. In Refs.
\cite{Enq3}-\cite{Once} the fermion contributions were not taken
into consideration. However, at high temperature they affect the
vacuum considerably. Quark posses both the electric and the color
charges and therefore the quark loops change the strengths of the
simultaneously generated magnetic and chromomagnetic fields \cite{DS}.

In a supersymmetric theory new peculiarities should be accounted
for. First is an influence of superpartners. These particles
having a low spin have to decrease the generated magnetic field
strengths. Second, s-quarks also possess the electric and the
color charges, so the interdependence of magnetic and
chromomagnetic fields is expected to be stronger as well as the
fields due to their vacuum loops. Because of this mixing some
specific configurations of the fields must be produced at high
temperature.

In the present paper the spontaneous vacuum magnetization is
investigated in the minimal supersymmetric standard model (MSSM)
of elementary particles. All boson and fermion fields are taken
into consideration. In the MSSM there are two kinds of non-abelian
gauge fields - the $SU(2)$ weak isospin gauge fields responsible
for weak interactions and the $SU(3)$ gluons mediating the strong
interactions. Magnetic and chromomagnetic fields are related to
these symmetry groups, respectively. To elaborate this problem we
calculate the effective potential (EP) including the one-loop and
the daisy diagram contributions in the constant abelian
chromomagnetic and magnetic fields, $H_c=const$ and $H=const$, at
high temperatures. The values of the generated fields strengths
are found as the minimum position of the EP in the field strength
plane $(H,H_c)$.

Let us note the advantages of the used approximation. The EP of
the background abelian magnetic fields is a gauge fixing
independent one, while the daisy diagrams account for the most
essential long-range corrections at high temperature. Therefore,
such a type of EP includes the leading and the next-to-leading
terms in the coupling constants. Moreover, as it was shown in
\cite{Once,Gio}, the daisy diagrams of the charged gluons and the
W-bosons with their magnetic masses taken into consideration make
the vacuum with nonzero magnetic fields stable at high
temperatures. The stability reflects the consistency of the
approximation. The EP of this type was used recently in
investigations of the electroweak phase transition in an external
hypercharge magnetic field \cite{UFZ} and the spontaneous
generation of magnetic and chromomagnetic fields in the SM
\cite{DS}. The obtained results are in a good agreement with the
non-perturbative calculations carried out in \cite{Kaj,Lai}.
This approximation will be used in what follows.

The abelian hypercharge magnetic field is not generated
spontaneously. So, in our investigation we shall consider the
non-abelian constituent of the magnetic field coming from $SU(2)$
gauge group. The generation mechanisms of the hypermagnetic field
were studied in \cite{Gio,Joy}. Below it will be shown that at
high temperatures either strong magnetic or the chromomagnetic
fields are generated in the MSSM, similarly to the SM. The additional
to the SM sector of the MSSM, s-particle sector, does not suppress
this effect. It just decreases the strengths of generated fields.
These fields are stable in the approximation adopted due to the
magnetic masses $m_{transversal}^2\sim (gH)^{1/2} T$ of the gauge
field transversal modes \cite{SkSt}. In this way the consistent
picture of the magnetized vacuum state is derived.

The contents of this paper are as follows. In Sect. 2 the
contributions of all bosons and fermions to the EP $v^\prime(H,T)$
of external magnetic and chromomagnetic fields are calculated in a
form convenient for numeric investigations. In Sect. 3 the field
strengths are calculated. Discussion and concluding remarks are
given in Sect. 4.

\subsection*{2. Basic Formulae}
The full Lagrangian of the MSSM can be written as (\cite{Compl})
\begin{eqnarray}
{\cal L}&=&L_{gs}+L_{\bar{gs}}+L_{leptons}+L_{sleptons}
+L_{quarks}+L_{squarks}\\\nonumber
&&+L_{Higgs}+L_{higgsino}+L_{int}+L_{SSB}+L_{gf}.
\end{eqnarray}
Here, $L_{gs}$ and $L_{\bar{gs}}$ is the kinetic part of gauge
bosons and gauginos, correspondingly; $L_{leptons}$,
$L_{sleptons}$, $L_{quarks}$ and $L_{squarks}$ give the kinetic
part of the matter (fermions and s-fermions) fields and their
interaction Lagrangians; $L_{Higgs}$ is the kinetic part and Higgs
interactions with gauge bosons and gauginos; $L_{int}$ contains
the interaction terms; $L_{SSB}$ is the soft-symmetry breaking
(SSB) part; $L_{gf}$ is the gauge fixing terms.

In the MSSM there are mixings in the higgsino and gaugino sector
resulting in the presence of chargino (mixing of charged higgsino
and gaugino) and neutralino (mixing of neutral higgsino and
gaugino) in the theory. Since the electrical and color neutral
particles are not interacting with magnetic and chromomagnetic
fields, we will not take the neutralino into consideration.

The MSSM Lagrangian of the gauge boson sector is (see \cite{Compl})
\begin{eqnarray}
L_{gs}=-\frac14 F^\alpha_{\mu\nu}F^{\mu\nu}_\alpha-\frac14
G_{\mu\nu}G^{\mu\nu}-\frac14{\bf F}^\alpha_{\mu\nu}{\bf
F}_\alpha^{\mu\nu},
\end{eqnarray}
where the standard notation is introduced
\begin{eqnarray}
F^\alpha_{\mu\nu}=\partial_\mu A_\nu^\alpha-\partial_\nu
A^\alpha_\mu+g \epsilon^{abc}A_\mu^b A_\nu^c,\\\nonumber
G_{\mu\nu}=\partial_\mu B_\nu-\partial_\nu B_\mu,\\\nonumber
{\bf F}^\alpha_{\mu\nu}=\partial_\mu {\bf A}_\nu^\alpha-
\partial_\nu {\bf A}_\mu^\alpha+g_sf^{abc}{\bf A}^b_\mu {\bf
A}^c_\nu.
\end{eqnarray}
The fields corresponding to the gauge $W-$, $Z-$bosons and
photons, respectively, are
\begin{eqnarray}
W^\pm_\mu=\frac{1}{\sqrt{2}}(A^1_\mu\pm iA^2_\mu),\\\nonumber
Z_\mu=\frac{1}{\sqrt{g^2+g^{\prime 2}}} (gA^3_\mu-
g^\prime B_\mu),\\\nonumber
A_\mu=\frac{1}{\sqrt{g^\prime A^3_\mu+gB_\mu}}
\end{eqnarray}
and ${\bf A}^\alpha_\mu$ is gluon field.

As usually, the introduction of gauge fields is being done by
replacing all derivatives in the Lagrangian with the covariant
ones,
\begin{eqnarray}
\partial_\mu \to {\cal D}_\mu=\partial_\mu+ig\frac{\tau^\alpha}{2}
A^\alpha_\mu+ig_s\frac{\lambda^\alpha}{2}{\bf A}_\mu^\alpha.
\end{eqnarray}
Here $\tau^\alpha$ and $\lambda^\alpha$ stand for the Pauli and
the Gell-Mann matrices, respectively.

In the $SU(2)$ sector there is only one magnetic field, the third
projection of the gauge field. In the $SU(3)_c$ sector there are
two possible chromomagnetic fields connected with the third and
the eighth generators of the group.

For simplicity, in what follows we shall consider the field
associated with the third generator of the $SU(3)_c$.

To introduce the interaction with classical magnetic and
chromomagnetic fields we split the potentials in two parts:
\begin{eqnarray}
A_\mu=\bar{A}_\mu+A^R_\mu,\\\nonumber
{\bf A}_\mu=\bar{\bf A}_\mu+{\bf A}^R_\mu,
\end{eqnarray}
where $A^R$ and ${\bf A}^R$ describe the radiation fields and
$\bar{A}=(0,0,Hx^1,0)$ and $\bar{\bf A}=(0,0,{\bf H}_3x^1,0)$
correspond to the constant magnetic and chromomagnetic fields
directed along the third axes in the space and in the internal
color and isospin spaces.

To construct the total EP we used the general relativistic
renormalizable gauge which is set by the following gauge fixing
conditions \cite{VVk}:
\begin{eqnarray}
\label{gauge1}
\partial_\mu W^{\pm \mu}\pm ie\bar{A}_\mu W^{\pm \mu} \mp
i \frac{g\phi_c}{2\xi}\phi^\pm=C^\pm(x),\\
\nonumber
\partial_\mu Z^\mu-\frac{i}{\xi^\prime}(g^2+g^{\prime
2})^{1/2}\phi_c \phi_Z=C^Z(x),\\
\nonumber
\partial_\mu {\bf A}^\mu + ig_s \bar{\bf A}={\bf C}(x),
\end{eqnarray}
where $e=g~sin \theta_W$, $tan\theta_W=g^\prime/g$, $\phi^\pm$ and
$\phi_Z$ are the Goldstone fields, $\xi$ and $\xi^\prime$ are the
gauge fixing parameters, $C^\pm$ and $C^Z$ are arbitrary functions
and $\phi_c$ is the value of the scalar field condensate. Setting
$\xi,\xi^\prime=0$ we choose the unitary gauge. In the restored phase
the scalar field condensate $\phi_c=0$ and the equations (\ref{gauge1})
are simplified.

The values of the macroscopic magnetic and chromomagnetic fields
generated at high temperature will be calculated by minimization
of the thermodynamics potential $\Omega$ which is introduced as follows
\begin{eqnarray}
\label{Omeg}
\Omega=-\frac{1}{\beta}log~Z,\\
Z=Tr~exp(-\beta{\cal{H}}),
\end{eqnarray}
where $Z$ is the partition function, and ${\cal{H}}$ is the
Hamiltonian of the system. The trace is calculated over all
physical states.

To obtain the EP one has to rewrite (\ref{Omeg}) as a sum in
quantum states calculated near the nontrivial classical solutions
$A^{ext}$ and ${\bf A}^{ext}$. This procedure is well-described
in the literature (see, for instance, \cite{Once,Kap,Carr})
and the result can be written in the form
\begin{eqnarray}
V=V^{(1)}(H,{\bf H}_3,T)+V^{(2)}(H,{\bf H}_3,T)+...
+V_{daisy}(H,{\bf H}_3,T)+...,
\end{eqnarray}
where $V^{(1)}$ is the one-loop EP; the other terms present the
contributions of two-, three-, etc. loop corrections.

Among these terms there are some responsible for the dominant
contributions of long distances at high temperature - so-called
daisy or ring diagrams (see, for example, \cite{Kap}). This part
of the EP, $V_{daisy}(H,{\bf H}_3,T)$, is nonzero in case when
massless states appear in a system. The ring diagrams have to be
calculated when the vacuum magnetization at finite temperature is
investigated. In fact, one first must assume that the fields are
nonzero, calculate EP $V(H,{\bf H}_3,T)$ and after that check
whether its minimum is located at nonzero $H$ and ${\bf H}_3$. On
the other hand, if one investigates problems in the applied
external fields, the charged fields become massive with the masses
depending on $\sim(gH)^{1/2}$, $\sim(g_s{\bf H}_3)^{1/2}$  and
have to be omitted.

The one-loop contribution to EP is given by the expression
\begin{equation}
\label{Trace}
V^{(1)}=-\frac12 Tr~log~G^{ab},
\end{equation}
where $G^{ab}$ stands for the propagators of all quantum
fields $W^\pm$, ${\bf A}$, $\ldots$ in the background fields $H$
and ${\bf H}_3$. In the proper time formalism, $s$-representation,
the calculation of the trace
can be carried out in accordance with the formula \cite{Schw}
\begin{equation}
Tr~log~G^{ab}=-\int_0^\infty\frac{ds}{s}tr~exp(-isG^{-1}_{ab}).
\end{equation}
Details of calculations based on the $s$-representation and
formula (\ref{Green}) can be found in \cite{Cab,Rez,Ska}.

To incorporate the temperature into this formalism in a natural
way we make use the method of \cite{Cab} which connects the
Green functions at zero temperature with the Matsubara Green
functions,
\begin{eqnarray}
\label{Green}
G^{ab}_k(x,x^\prime;T)=\sum^{+\infty}_{-\infty}
(-1)^{(n+[x])\sigma_k}G^{ab}_k(x-[x]\beta u,x^\prime-n\beta u),
\end{eqnarray}
where $G^{ab}_k$ is the corresponding function at $T=0$,
$\beta=1/T$, $u=(0,0,0,1)$, $[x]$ denotes an integer
part of $x_4/\beta$, $\sigma_k=1$ in the case of physical fermions
and $\sigma_k=0$ for boson and ghost fields. The Green functions
in the right-hand side of (\ref{Green}) are the matrix elements of
the operators $G_k$ computed in the states $\vert
x^\prime,a\rangle$ at $T=0$, and in the left-hand side the
operators are averaged over the states with $T\neq 0$. The
corresponding functional spaces $U^0$ and $U^T$ are different but
in the limit of $T\to 0$ $U^T$ transforms into $U^0$.

The terms with $n=0$ in (\ref{Green}) and (\ref{Trace}) give the
zero temperature expressions for the Green functions and the EP
$V^\prime$, respectively. So, we can split the latter into two
parts:
\begin{equation}
V^\prime(H,{\bf H}_3,T)=V^\prime(H,{\bf H}_3)
+V^\prime_\tau (H,{\bf H}_3,T).
\end{equation}
The standard procedure to account for the daisy diagrams is to
substitute the tree level Matsubara Green functions in
(\ref{Trace}) $[G^{(0)}_i]^{-1}$ by the  full propagator
$G^{-1}_i=[G^{(0)}_i]^{-1}+\Pi(H,T)$ (see for details
\cite{Once,Kap,Carr}), where the last term is the polarization
operator at finite temperature in the field taken at zero
longitudinal momentum $k_l=0$.

Omitting the detailed calculations we notice that the exact
one-loop EP is transformed into the EP which contains the daisy
diagrams as well as the one-loop diagrams if one adds to the
exponent a term containing the temperature dependent mass of
a particle.

It is convenient for what follows to introduce the dimensionless
quantities: $x=H/H_0$ $(H_0=M_W^2/e)$, $y={\bf H}_3/{\bf H}_3^0$
$({\bf H}_3^0=M_W^2/g_s)$, $B=\beta M_W$, $\tau=1/B=T/M_W$, $v=V/H_0^2$.

The total EP consists of several terms
\begin{eqnarray}
\label{vprime}
v^\prime=\frac{x^2}{2}+\frac{y^2}{2}&+&v^\prime_{leptons}+v^\prime_{quarks}
+v^\prime_{W-bosons}+v^\prime_{gluons}
\\\nonumber&&+v^\prime_{sleptons}+v^\prime_{squarks}+v^\prime_{charginos}
+v^\prime_{gluinos}.
\end{eqnarray}

These terms can be written down as follows (in dimensionless variables):

\vskip 0.5 cm
$\bullet$ {\bf SM sector}
\vskip 0.5 cm

{\sl - leptons}
\begin{eqnarray}
v^\prime_{leptons}=-\frac{1}{4 \pi^2}\sum_{n=1}^{\infty}
(-1)^n \int_0^\infty\frac{ds}{s^3}
e^{-(m_{leptons}^2 s+\frac{\beta^2n^2}{4s})}(xs~Coth(xs)-1);
\end{eqnarray}

{\sl - quarks}
\begin{eqnarray}
v^\prime_{quarks}=-\frac{1}{4 \pi^2}\sum_{f=1}^6\sum_{n=1}^{\infty}
(-1)^n \int_0^\infty
\frac{ds}{s^3}e^{-(m_{f}^2 s+\frac{\beta^2 n^2}{4s})}\\\nonumber
\cdot(q_fxs~Coth(q_fxs)\cdot ys~Coth(ys)-1);
\end{eqnarray}

{\sl - $W$-bosons} (see \cite{EW})
\begin{eqnarray}
v^\prime_W=-\frac{x}{8\pi^2}\sum_{n=1}^\infty
\int_0^\infty\frac{ds}{s^2}e^{-(m_{W}^2 s+\frac{\beta^2
n^2}{4s})}\cdot\left[\frac{3}{Sinh(xs)}+4Sinh(xs)\right];
\end{eqnarray}

{\sl - gluons} (see \cite{Once})
\begin{eqnarray}
\label{gluons}
v^\prime_{gluons}=-\frac{y}{4\pi^2}\sum_{n=1}^\infty
\int_0^\infty\frac{ds}{s^2}e^{-(m_{gluons}^2 s+\frac{\beta^2 n^2}{4s})}
\cdot\left[\frac{1}{Sinh(ys)}+2Sinh(ys)\right];
\end{eqnarray}

\vskip 0.5 cm
$\bullet$ {\bf MSSM sector}
\vskip 0.5 cm

{\sl - s-leptons}
\begin{eqnarray}
v^\prime_{sleptons}=-\frac{3}{4\pi^2}\sum_{n=1}^\infty
\int_0^\infty\frac{ds}{s^3}e^{-(m_{sleptons}^2 s+\frac{\beta^2
n^2}{4s})}\left[\frac{xs}{Sinh(xs)}-1\right];
\end{eqnarray}

{\sl - s-quarks}
\begin{eqnarray}
v^\prime_{squarks}=-\frac{1}{8\pi^2}\sum_{n=1}^\infty
\int_0^\infty\frac{ds}{s^3}e^{-(m_{squarks}^2 s+\frac{\beta^2
n^2}{4s})}\left[\frac{q_fxs\cdot ys}{Sinh(q_fxs)\cdot Sinh(ys)}-1\right];
\end{eqnarray}

{\sl - charginos}
\begin{eqnarray}
v^\prime_{charginos}=-\frac{1}{4 \pi^2}\sum_{n=1}^{\infty}
(-1)^n \int_0^\infty\frac{ds}{s^3}
e^{-(m_{charginos}^2 s+\frac{\beta^2n^2}{4s})}(xs~Coth(xs)-1);
\end{eqnarray}

{\sl - gluinos}
\begin{eqnarray}
\label{gluinos}
v^\prime_{gluinos}=-\frac{1}{4 \pi^2}\sum_{n=1}^{\infty}
(-1)^n \int_0^\infty\frac{ds}{s^3}
e^{-(m_{gluinos}^2 s+\frac{\beta^2n^2}{4s})}(ys~Coth(ys)-1).
\end{eqnarray}

Here, $m_{leptons}$, $m_{f}$, $m_{W}$, $m_{gluons}$,
$m_{sleponts}$, $m_{squarks}$, $m_{charginos}$ and $m_{gluinos}$
are the temperature masses of leptons, quarks, W-bosons, gluons,
s-leptons, s-quarks, charginos and gluinos, respectively;
$q_f=\left(\frac23,-\frac13,-\frac13,\frac23,-\frac13,\frac23\right)$
are the charges of quarks.

Since we investigate the high-temperature effects
connected with the presence of external fields, we used
leading in temperature terms of the Debye masses of the particles,
only (\cite{Once,EW}).

In present analysis the temperature masses of leptons, quarks,
s-leptons, s-quarks, charginos, gluinos are taken as follows
(\cite{DS})
\begin{eqnarray}
\begin{array}{ll}
m_{leptons}^2=\left(\frac{e}{\beta}\right)^2, &
m_{f}^2=\left(\frac{e}{\beta}\right)^2,\\
m_{sleptons}^2=\left(\frac{e}{\beta}+\frac{M_{sleptons}}{M_W}\right)^2, &
m_{squarks}^2=\left(\frac{e}{\beta}+\frac{M_{squarks}}{M_W}\right)^2,\\
m_{charginos}^2=\left(\frac{e}{\beta}+\frac{M_{charginos}}{M_W}\right)^2, &
m_{gluinos}^2=\left(\frac{e}{\beta}+\frac{M_{gluinos}}{M_W}\right)^2,
\end{array}
\end{eqnarray}
where the masses from SSB terms are taken as the low experimental
limits of corresponding particle masses (\cite{PDG})
\begin{eqnarray}
\begin{array}{ll}
M_{sleptons}=40~GeV, & M_{squarks}=176~GeV,\\
M_{charginos}=62~GeV, & M_{gluinos}=154~GeV.
\end{array}
\end{eqnarray}

As it was established in numeric computation the spontaneous
generation of fields depends on SSB masses fairly weak. Even in
the case of zero SSB masses there is the generation of magnetic
and chromomagnetic fields. In the limit of infinite SSB masses the
picture conforms to the SM case.

The temperature masses of gluons and $W$-bosons are
(\cite{Once,DS,SkSt})
\begin{eqnarray}
\begin{array}{cc}
m_{W}^2=15~\alpha_{EW}~\frac{x^{1/2}}{\beta}, &
m_{gluons}^2=15~\alpha_S~\frac{y^{1/2}}{\beta},
\end{array}
\end{eqnarray}
$\alpha_{EW}$ and $\alpha_S$ are the electroweak and
the strong interaction couplings, respectively.

In one-loop order, the neutral gluon contribution is trivial
${\bf H}_3-$independent constant which can be omitted. However,
these fields are long-range states and they do give
${\bf H}_3-$dependent EP through the correlation corrections
depending on the temperature and field. We include the
longitudinal neutral modes only because their Debye masses
$\Pi^0(y,\beta)$ are nonzero. The corresponding EP is
\cite{Once}
\begin{eqnarray}
\label{ring}
v_{ring}=\frac{1}{24\beta^2}\Pi^0(y,\beta)-\frac{1}{12\pi\beta}
\left(\Pi^0(y,\beta)\right)^{3/2}\\\nonumber
+\frac{\left(\Pi^0(y,\beta)\right)^2}{32\pi^2}
\left[log\left(\frac{4\pi}{\beta(\Pi^0(y,\beta))^{1/2}}\right)
+\frac{3}{4}-\gamma\right];
\end{eqnarray}
$\gamma$ is Euler's constant,
$\Pi^0(y,\beta)=\Pi^0_{00}(k=0,y,\beta)$ is the zero-zero
component of the neutral gluon field polarization operator
calculated in the external field at finite temperature and taken
at zero momentum \cite{Once}
\begin{eqnarray}
\Pi^0(y,\beta)=\frac{2 g^2}{3
\beta^2}-\frac{y^{1/2}}{\pi\beta}-\frac{y}{4\pi^2}.
\end{eqnarray}

Equations (\ref{vprime})-(\ref{gluinos}) and (\ref{ring}) will be
used in numeric calculations.

\subsection*{3. Combined generation of magnetic and chromomagnetic fields}
To calculate the strengths of combined generated magnetic and
chromomagnetic fields we use the perturbative computation method
in \cite{DS}. First of all we find the strengths of the fields $x$
and $y$ when the quark and the s-quark contributions
($v^\prime_q$) are divided in two parts,
$v^\prime_q(x,\beta)=v^\prime_q\mid_{y\to 0}$ and
$v^\prime_q(y,\beta)=v^\prime_q\mid_{x\to 0}$, where $v^\prime_q
(x,\beta)$ is the quark and s-quark contribution in the case of single
magnetic field, and $v^\prime_q (y,\beta)$ is the one in the
presence of chromomagnetic field, only.

Now let us rewrite $v^\prime$ in (\ref{vprime}) as follows:
\begin{equation}
v^\prime(\bar{x},\bar{y})=v_1(\bar{x})+v_2(\bar{y})+v_3(\bar{x},\bar{y}),
\end{equation}
where $\bar{x}=x+\delta x$, $\bar{y}=y+\delta y$, and $\delta x$
and $\delta y$ are the field corrections connected with the
interfusion effect of the fields in the quark and s-quark
sector.

Since the mixing of fields due to quark and s-quark loop is weak
(this is justified in numeric calculations) one can assume
that $\delta x \ll 1$ and $\delta y \ll 1$, and write
\begin{eqnarray}
v_1(\bar{x})=v_1(x+\delta x)=v_1(x)+\frac{\partial v_1 (x)}{\partial
x}\delta x,\\
v_2(\bar{y})=v_2(y+\delta y)=v_2(y)+\frac{\partial v_2 (y)}{\partial
y}\delta y,\\
v_3(\bar{x},\bar{y})=v_3(x+\delta x,y+\delta y)=v_3(x,y).
\end{eqnarray}
After simple transformations we can find $\delta x$ and
$\delta y$:
\begin{eqnarray}
\delta x=\frac{\frac{\partial v_3(x,0)}{\partial x}-
\frac{\partial v_3(x,y)}{\partial x}}
{\frac{\partial^2 v_1(x)}{\partial x^2}},\nonumber\\
\delta y=\frac{\frac{\partial v_3(0,y)}{\partial y}-
\frac{\partial v_3(x,y)}{\partial y}}
{\frac{\partial^2 v_2(y)}{\partial y^2}}.
\end{eqnarray}

Hence we obtain $\bar{x}=x+\delta x$ and $\bar{y}=y+\delta y$.

These results on the field strengths determined by means of numeric
investigation of the total EP are summarized in Tables 1 and 2.

\vskip 0.5 cm
\begin{center}
\begin{tabular}{|c|l|c|c|l|l|}
\hline
\multicolumn{1}{|c|}{ } & \multicolumn{4}{|c|}{\bf MSSM} & \multicolumn{1}{|c|}{\bf SM} \\
\cline{2-6}
$\beta$ & \multicolumn{1}{|c|}{$x$} & $\delta x$ & $\delta x/x$, \% &
\multicolumn{1}{|c|}{$\bar{x}$} & \multicolumn{1}{|c|}{$\bar{x}$}
\\\hline\hline
$0.1$ & $0.3813$    & $1.58\times 10^{-2}$ & $4.14$ & $0.3971$    & $0.7000$ \\
$0.2$ & $0.10021$   & $2.45\times 10^{-3}$ & $2.44$ & $0.10265$   & $0.20075$ \\
$0.3$ & $0.046199$  & $7.19\times 10^{-4}$ & $1.56$ & $0.046917$  & $0.069945$ \\
$0.4$ & $0.026804$  & $1.97\times 10^{-4}$ & $0.73$ & $0.027000$  & $0.039964$ \\
$0.5$ & $0.017675$  & $1.19\times 10^{-4}$ & $0.67$ & $0.017794$  & $0.029953$ \\
$0.6$ & $0.0120559$ & $5.20\times 10^{-5}$ & $0.43$ & $0.0121079$ & $0.0199508$ \\
$0.7$ & $0.0086022$ & $2.82\times 10^{-5}$ & $0.33$ & $0.0086303$ & $0.0099620$ \\
$0.8$ & $0.0065687$ & $1.72\times 10^{-5}$ & $0.26$ & $0.0065859$ & $0.0099381$ \\
$0.9$ & $0.0052535$ & $1.13\times 10^{-5}$ & $0.22$ & $0.0052648$ & $0.0099759$ \\
$1.0$ & $0.0043400$ & $8.10\times 10^{-6}$ & $0.19$ & $0.0043481$ & $0.0099643$\\
\hline
\end{tabular}
\end{center}
\medskip

\noindent{{\bf Table 1.} {\it The strengths of generated magnetic field.}}\par
\vskip 0.5 cm

In Tables 1 and 2, in the first column we show the inverse
temperature. In the second one the strengths of magnetic and
chromomagnetic fields are adduced for the case of the quark and the
s-quark EP describing each of the fields separately. The next column
gives the field corrections in the case of the total quark and s-quark
EP. The fourth column presents the relative value of corrections.
The following column gives the resulting strengths of magnetic
($\bar{x}=x+\delta x$) and chromomagnetic ($\bar{y}=y+\delta y$)
fields, respectively. In the last column the strengths of
generated fields in the SM are given for comparison (\cite{DS}).

As it is seen, the increase of inverse temperature leads to
decreasing the strengths of generated fields. This dependence is
well in accordance with the picture of  the universe cooling.

\vskip 0.5 cm
\begin{center}
\begin{tabular}{|c|l|c|c|l|l|}
\hline
\multicolumn{1}{|c|}{ } & \multicolumn{4}{|c|}{\bf MSSM} & \multicolumn{1}{|c|}{\bf SM} \\
\cline{2-6}
$\beta$ & \multicolumn{1}{|c|}{$y$} & $\delta y$ & $\delta y/y$, \% &
\multicolumn{1}{|c|}{$\bar{y}$} & \multicolumn{1}{|c|}{$\bar{y}$}
\\\hline\hline
$0.1$ & $0.510146$   & $5.28\times 10^{-6}$ & $0.0010$ & $0.510151$   & $0.800301$ \\
$0.2$ & $0.133035$   & $1.73\times 10^{-6}$ & $0.0013$ & $0.133037$   & $0.199761$ \\
$0.3$ & $0.0603172$  & $8.85\times 10^{-7}$ & $0.0015$ & $0.0603181$  & $0.0899012$ \\
$0.4$ & $0.0347127$  & $5.20\times 10^{-7}$ & $0.0015$ & $0.0347132$  & $0.0499116$ \\
$0.5$ & $0.0225367$  & $3.59\times 10^{-7}$ & $0.0016$ & $0.0225371$  & $0.0398880$ \\
$0.6$ & $0.0161563$  & $2.26\times 10^{-7}$ & $0.0014$ & $0.0161565$  & $0.0299018$ \\
$0.7$ & $0.0115808$  & $1.53\times 10^{-7}$ & $0.0013$ & $0.0115810$  & $0.0199558$ \\
$0.8$ & $0.00859328$ & $1.19\times 10^{-7}$ & $0.0014$ & $0.00859340$ & $0.0199267$ \\
$0.9$ & $0.00672412$ & $9.94\times 10^{-8}$ & $0.0015$ & $0.00672422$ & $0.0098830$ \\
$1.0$ & $0.00547797$ & $9.01\times 10^{-8}$ & $0.0016$ & $0.00547806$ & $0.0098250$\\
\hline
\end{tabular}
\end{center}
\medskip

\noindent{{\bf Table 2.} {\it The strengths of generated chromomagnetic field.}}\par
\vskip 0.5 cm

From the above analysis it follows that in the considered temperature
interval the presence in the system of both fields leads to
increasing of each of them in contrast with the SM case. In the
latter the strengths of the combined fields are decreased as
compared to the separate generation. This is the consequence of
the s-quark loop contributions depending as the quark loops on
both of fields.

With temperature decreasing this effect becomes less pronounced
and disappears at comparably low temperatures $\beta\sim 1$.

\begin{figure}
\begin{center}
\resizebox{0.85\textwidth}{!}{\includegraphics{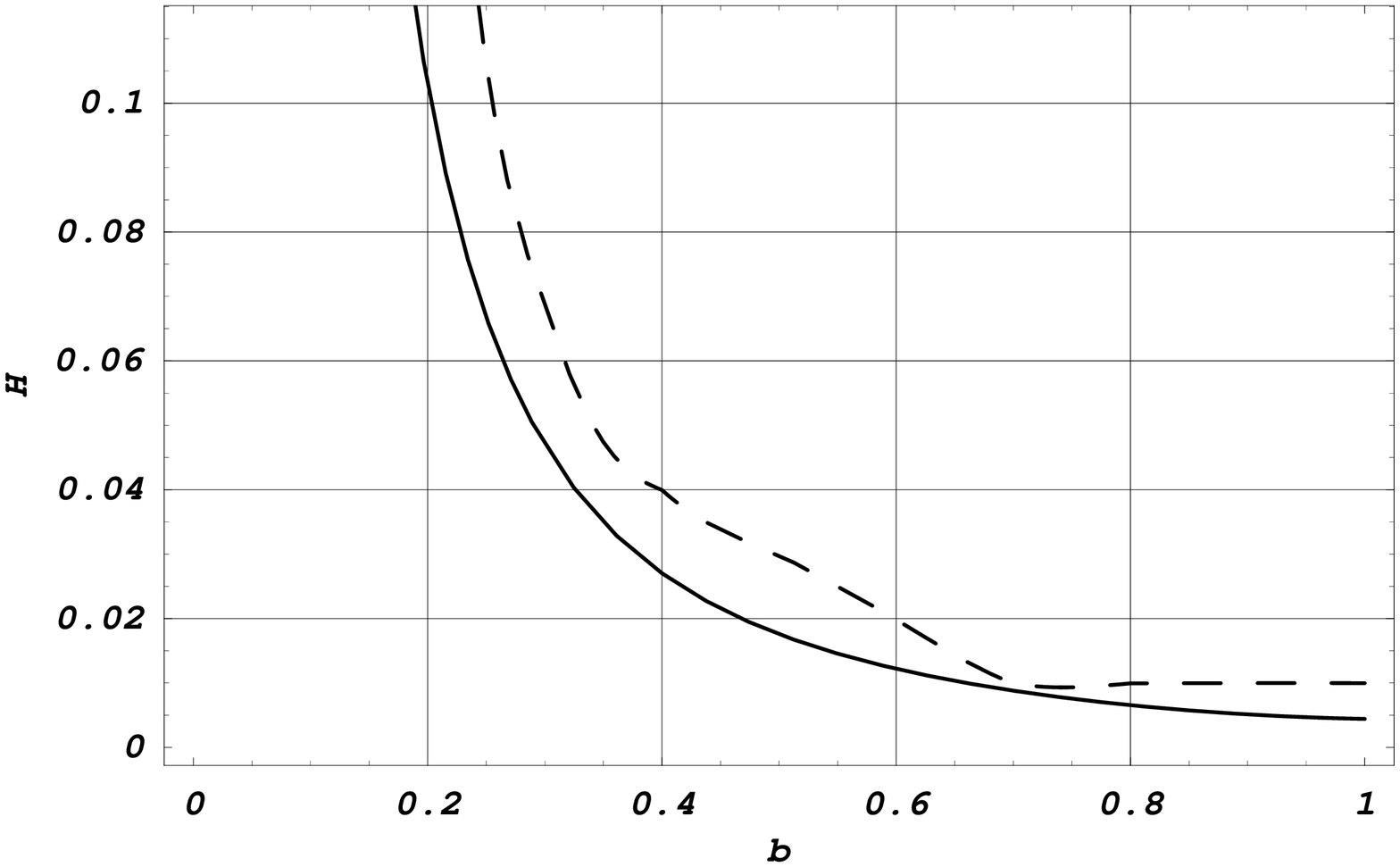}}
\caption{The dependences of the strengths of generated magnetic
fields (${\bf H}$) on the inverse temperature (${\bf b}$). The solid line is the
magnetic field strength in the MSSM and the dashed line is that of in the SM.}
\label{fig:1}
\end{center}
\end{figure}

\begin{figure}
\begin{center}
\resizebox{0.85\textwidth}{!}{\includegraphics{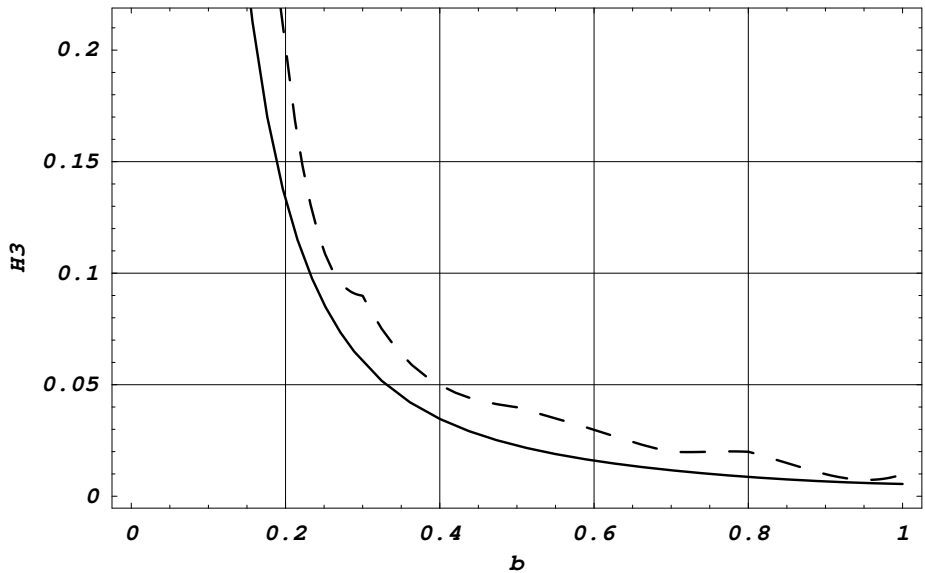}}
\caption{The dependences of the strengths of generated chromomagnetic
field (${\bf H}$) on the inverse temperature (${\bf b}$). The solid line is the
chromomagnetic field strength in the MSSM and the dashed line is that of in
the SM.}
\label{fig:2}
\end{center}
\end{figure}

\subsection*{4. Discussion}
Let us discuss the results obtained. As we elaborated within the
EP including the one-loop and the daisy diagrams, in the MSSM at
high temperatures both the magnetic and chromomagnetic fields have
to be generated. This vacuum is stable, as it follows from the
absence of imaginary terms in the EP minima.

If quark and s-quark loops are discarded, both of the fields can
be generated separately. All these states are stable, due to the
magnetic mass $\sim g^2(gH)^{1/2} T$ of the transversal gauge
field modes. As it was already shown in \cite{DS}, the imaginary
part arises for the field strengths larger than the ones generated
in the SM. As we have seen, the strengths of generated fields
reduce due to the s-particle sector of the MSSM. This vacuum state
is more stable as compare to the SM case.

The result on the stabilization of the charged gauge field spectra
is very important. It has relevance not only to the problem of the
consistent description of the generation of magnetic fields but
also to the related problem on symmetry behavior in external
magnetic fields investigated in the MSSM recently in
\cite{Gio,Joy}.

As it is seen from Figs. 1 and 2, presenting the results of numeric
computations within the exact EP, the strengths of the generated
fields are increasing when the temperature is rising. It is also
found, the dynamics of curves obtained in the SM \cite{DS} are
in a good agreement with our numeric calculations.

The ground state possessing the magnetic and the chromomagnetic
fields makes it reasonable to expect the existence of these fields
in the electroweak transition epoch for both the SM and the
MSSM. The state is stable in the whole considered temperature
interval. The imaginary part in the EP exists for the external
fields much stronger than the strengths of the spontaneously
generated ones. The mixing of magnetic and chromomagnetic fields
arising from the quark and the s-quark sectors of the EP is weak.
In the MSSM, the change of the field minima in the inclusion of
the field mixing does not exceed 4 per cents. In the SM these
values do not exceed 2 per cents. This is due to the strong
dependence of the s-quark loop on the strengths of both fields.

\vskip 0.5 cm
\begin{center}
\begin{tabular}{|c|l|l|l|l|}
\hline
$\beta$ & \multicolumn{1}{|c|}{quarks} & \multicolumn{1}{|c|}{s-quarks}
& \multicolumn{1}{|c|}{s-leptons} & \multicolumn{1}{|c|}{charginos}\\
\hline\hline
$0.1$ & $8.09496\times 10^{-2}$ & $1.29860\times 10^{-2}$ & $1.08818\times 10^{-2}$ & $1.90295\times 10^{-3}$ \\
$0.2$ & $5.51654\times 10^{-3}$ & $5.13130\times 10^{-4}$ & $6.28152\times 10^{-4}$ & $1.16324\times 10^{-4}$ \\
$0.3$ & $1.13926\times 10^{-3}$ & $6.77157\times 10^{-5}$ & $1.13423\times 10^{-4}$ & $2.20561\times 10^{-5}$ \\
$0.4$ & $3.78190\times 10^{-4}$ & $1.51991\times 10^{-5}$ & $3.28547\times 10^{-5}$ & $6.66057\times 10^{-6}$ \\
$0.5$ & $1.60125\times 10^{-4}$ & $4.51358\times 10^{-6}$ & $1.24231\times 10^{-5}$ & $2.60925\times 10^{-6}$ \\
$0.6$ & $8.11555\times 10^{-5}$ & $1.64583\times 10^{-6}$ & $5.07024\times 10^{-6}$ & $1.09729\times 10^{-6}$ \\
$0.7$ & $4.16467\times 10^{-5}$ & $6.18990\times 10^{-7}$ & $2.28091\times 10^{-6}$ & $5.06211\times 10^{-7}$ \\
$0.8$ & $2.31108\times 10^{-5}$ & $2.55271\times 10^{-7}$ & $1.18256\times 10^{-6}$ & $2.68001\times 10^{-7}$ \\
$0.9$ & $1.42387\times 10^{-5}$ & $1.18144\times 10^{-7}$ & $6.76190\times 10^{-7}$ & $1.55895\times 10^{-7}$ \\
$1.0$ & $9.48895\times 10^{-6}$ & $5.96475\times 10^{-8}$ & $4.14475\times 10^{-7}$ & $9.68813\times 10^{-8}$ \\
\hline
\end{tabular}
\end{center}
\medskip

\noindent{{\bf Table 3.} {\it The contribution of
quarks, s-quarks, s-leptons and charginos to the EP.}}\par
\vskip 0.5 cm

During the universe cooling the strengths of the generated
fields are decreasing. This is in agreement with what is expected
in cosmology.

One of the consequences of the results obtained is the presence of
a strong chromomagnetic field in the early universe, in
particular, at the electroweak phase transition and, probably,
till the deconfinement phase transition. The influence of this
field on the phase transitions may bring new insight to these
phenomena. As our estimate showed, the chromomagnetic field is as
strong as the magnetic one. So the role of strong interactions in
the early universe in the presence of the field needs more
detailed investigations as compare to what is usually assumed
\cite{Grasso}.

We would like to notice that in the literature devoted to
investigations of the quark-gluon plasma in the deconfinement
phase carried out by non-perturbative methods the vacuum
magnetization at high temperature has not been accounted for (see,
for instance, recent paper \cite{Kaj2} and references therein).
From the point of view of the present analysis (as well as other
studies carried out already in perturbation theory
\cite{Enq3}-\cite{Once}) these investigations are incomplete. The
generation of the chromomagnetic field at high temperature has to
be taken into consideration.

\end{document}